\documentclass[12pt]{article}
\usepackage{amssymb}
\usepackage{times}
\usepackage{titling}
\usepackage{graphicx}
\usepackage{amsmath}
\usepackage[version=3]{mhchem}
\usepackage{color}
\usepackage{bm}
\usepackage{lineno}
\usepackage{multirow}
\providecommand{\tabularnewline}{\\}
\usepackage[paper=letterpaper,twoside,top=2.0cm,bottom=2cm,left=1.5cm,
            right=1.5cm,bindingoffset=0.0cm,voffset=0.0cm]{geometry}
\newcommand{\red}[1]{{#1}}

\title{
{\bf Continuous excitations of the triangular-lattice \\ quantum spin liquid YbMgGaO{$_4$}} \\}
\author{
 	Joseph A.~M. Paddison$^1$, Marcus Daum$^{1,\ddagger}$, Zhiling Dun$^{2,\ddagger}$, Georg Ehlers$^3$, \\ Yaohua Liu$^3$, Matthew B. Stone$^3$, Haidong Zhou$^2$, Martin Mourigal$^{1,\ast}$ \\
	\normalsize{$^1$School of Physics, Georgia Institute of Technology, Atlanta, GA 30332, USA } \\
	\normalsize{$^2$Department of Physics and Astronomy, University of Tennessee, Knoxville, TN 37996, USA} \\
	\normalsize{$^3$Quantum Condensed Matter Division, Oak Ridge National Laboratory, Oak Ridge, TN 37831, USA} \\
	\\
	\normalsize{\red{$^{\ddagger}$ These authors contributed equally to this work}} \\
	\normalsize{$^{\ast}$ Email: mourigal@gatech.edu} \\
	}
\date{July 12, 2016}

\begin{document}
\maketitle

%------------------------------------------
\baselineskip24pt

\textbf{A quantum spin liquid (QSL) is an exotic state of matter in which electrons' spins are quantum entangled over long distances, but do not show symmetry-breaking magnetic order in the zero-temperature limit \cite{Balents_2010}. The observation of QSL states is a central aim of experimental physics, because they host collective excitations that transcend our knowledge of quantum matter \cite{Anderson_1987,Lee_2008}; however, examples in real materials are scarce \cite{Lee_2008a}. Here, we report neutron-scattering measurements \red{on YbMgGaO{$_4$}, a QSL candidate} in which Yb$^{3+}$ ions with effective spin-$1/2$ occupy a triangular lattice \cite{Li_2015a,Li_2015}. Our measurements reveal a continuum of magnetic excitations---the \red{essential} experimental hallmark of a QSL \cite{Kalmeyer_1987,Han_2012}---at very low temperature ($\approx0.06$\,K). The origin of this \red{peculiar excitation spectrum} is a crucial question, because isotropic nearest-neighbor interactions do \emph{not} yield a QSL ground state on the triangular lattice \cite{Capriotti_1999}. Using measurements of the \red{magnetic excitations} close to the field-polarized state, we identify antiferromagnetic next-nearest-neighbor interactions \cite{Manuel_1999,Li_2015c,Zhu_2015,Hu_2015,Iqbal_2016} in the presence of planar anisotropy \cite{Li_2015} as key ingredients for QSL formation in YbMgGaO{$_4$}.}

\clearpage

%------------------------------------------
\baselineskip20pt

%================  ================  ================  ================
\begin{figure}
	\begin{center}
	\includegraphics[scale=0.70]{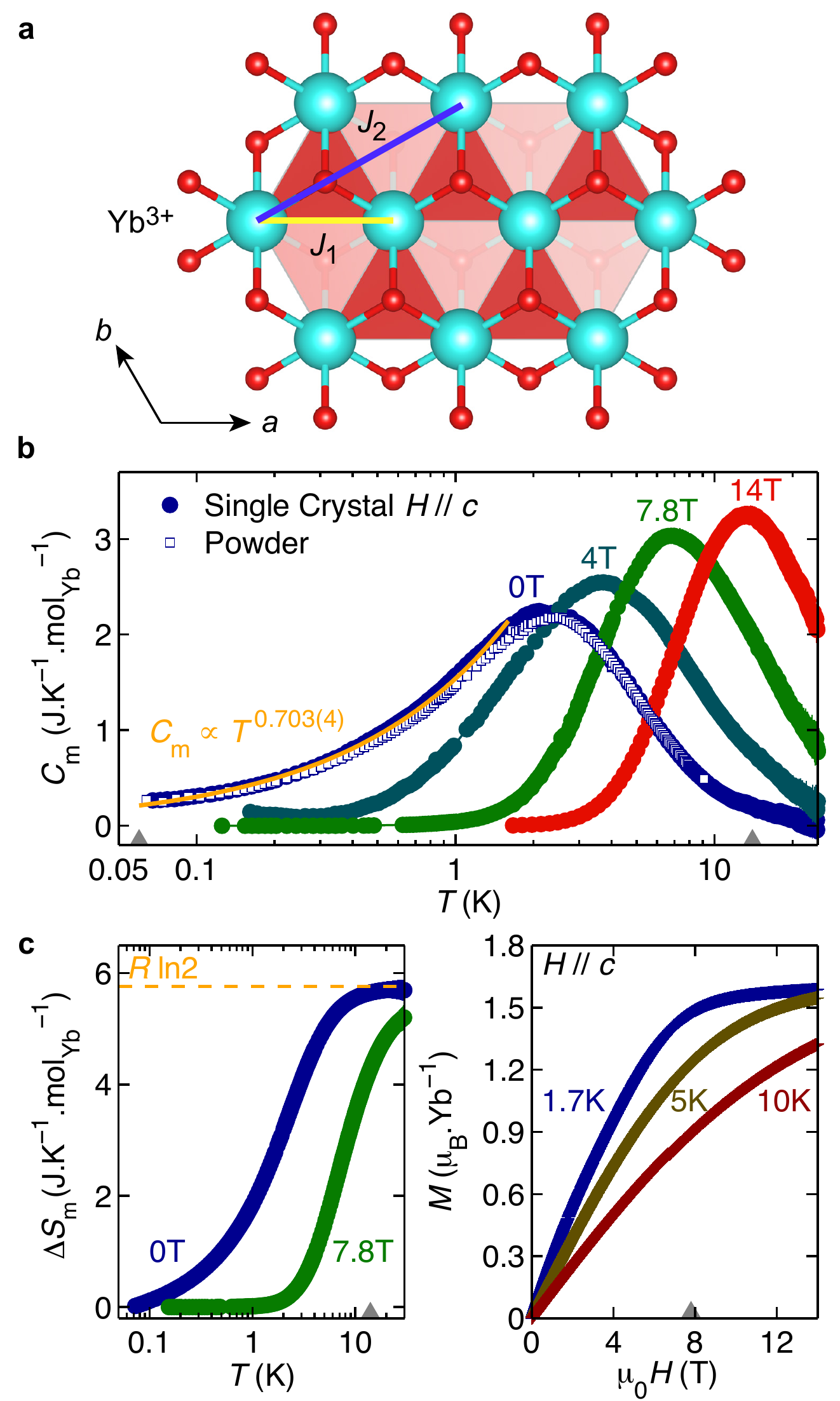}
	\end{center}
	\caption{\label{fig1}{\bf Structure and thermodynamic properties of YbMgGaO{$_4$}}. {\bf a.} Partial crystal structure, showing a triangular layer of Yb$^{3+}$ ions (large cyan spheres) and their coordination by oxygen (small red spheres). A nearest-neighbor interaction pathway $J_1$ and a next-nearest-neighbor interaction pathway $J_2$ are shown by yellow and blue lines, respectively. {\bf b.} Magnetic component of the specific heat $C_{\mathrm{m}}(T)$, showing data measured on a powder sample (hollow squares) and a single-crystal sample (filled circles). Single-crystal data are shown for applied magnetic fields along the $c$ axis of 0, 4, 7.8, and 14\,T (labelled above each curve). The orange line shows a fit of the zero-field single-crystal data to a power law, $C_{\mathrm{m}}(T)\propto T^{0.703(4)}$. {\bf c.} Magnetic entropy change $\Delta S_{\mathrm{m}}$, showing data measured in zero field and in a 7.8\,T field (labelled on each curve). {\bf d.} Dependence of the magnetization $M$ on applied field $\mu_0 H$, showing data measured at temperatures of 1.7, 5, and 10\,K (labelled on each curve). In {\bf b}, {\bf c} and {\bf d} the temperatures and applied fields at which we performed neutron-scattering measurements [Figs.~2 and 3] are indicated by small grey arrows.  }
\end{figure}
%================  ================  ================  ================

The phenomenon of entanglement is one of the fundamental results of quantum mechanics. Among its most extraordinary manifestations are quantum spin liquids, in which a macroscopic number of spins are entangled but do not show conventional magnetic order \cite{Balents_2010}. \red{The earliest examples of QSLs were found in quasi-one-dimensional systems such as Heisenberg spin chains \cite{Tennant_1993} and in the distorted triangular-lattice antiferromagnet Cs$_2$CuCl$_4$ \cite{Coldea_2001}}. The search for QSLs in two and three-dimensional systems has focused on frustrated magnets, in which the lattice geometry prevents all magnetic interactions from being satisfied simultaneously \cite{Balents_2010,Lee_2008a}. In two dimensions (2D), the prototype of a QSL is Anderson's ``resonating valence bond" model \cite{Anderson_1973}. Its ground state is a superposition of all possible tilings of dimers on the triangular lattice, where each dimer is built from a pair of spin-$1/2$; its excitations are unpaired spin-$1/2$, which are delocalized. The presence of delocalized excitations with fractional quantum numbers leads to a \red{necessary experimental signature} of a QSL: a magnetic excitation spectrum that is continuous \red{in energy} but structured in \red{momentum space} \cite{Kalmeyer_1987,Han_2012}. Such a spectrum has indeed been observed using neutron-scattering measurements of ZnCu$_3$(OH)$_6$Cl$_2$  (``herbertsmithite") \cite{Han_2012,Vries_2009}, in which spins occupy a kagome lattice. However, the \red{stabilization of a continuous excitation spectrum by quantum fluctuations} in a \red{structurally-perfect} triangular-lattice magnet---the scenario originally proposed by Anderson \cite{Anderson_1973}---has remained an open question.

Recent experiments have identified the triangular-lattice magnet YbMgGaO{$_4$} as an exciting QSL candidate \cite{Li_2015a,Li_2015}. The crystal structure (space group $R\bar{3}m$) contains undistorted triangular planes of magnetic Yb$^{3+}$, separated by two planes of site-disordered Mg$^{2+}$ and Ga$^{3+}$ [Fig.~1{\bf a}] \cite{Li_2015a}. Thermodynamic measurements show the absence of conventional magnetic order to $T=0.06$\,K, far below the Weiss temperature $\theta_{\mathrm{W}}\approx -4$\,K; moreover, the apparent absence of zero-point entropy indicates that the system essentially occupies a single quantum state at 0.06\,K \cite{Li_2015a}. The crystal-field ground state of Yb$^{3+}$ is a Kramers doublet separated by a large energy gap of 37~meV from the first excited state~\cite{Li_2015}; hence, an effective spin-$1/2$ description is appropriate at low temperature, as for ``quantum spin ice" Yb$_2$Ti$_2$O$_7$ \cite{Ross_2011}. Electron-spin resonance (ESR) measurements indicate that the nearest-neighbor magnetic interaction is anisotropic and depends on the bond orientation \cite{Li_2015}. This anisotropy provides one potential mechanism to stablilize a QSL ground state on the triangular lattice \cite{Li_2015b}. However, several alternative mechanisms for QSL behavior are known, including further-neighbor interactions \cite{Manuel_1999,Li_2015c,Zhu_2015,Hu_2015,Iqbal_2016} and multi-spin interactions \cite{Misguich_1999,Motrunich_2005}. Experiments to reveal the nature of the potential triangular-lattice QSL in YbMgGaO{$_4$} are therefore crucial.

%================  ================  ================  ================
\begin{figure}
	\begin{center}
		\includegraphics[scale=0.7]{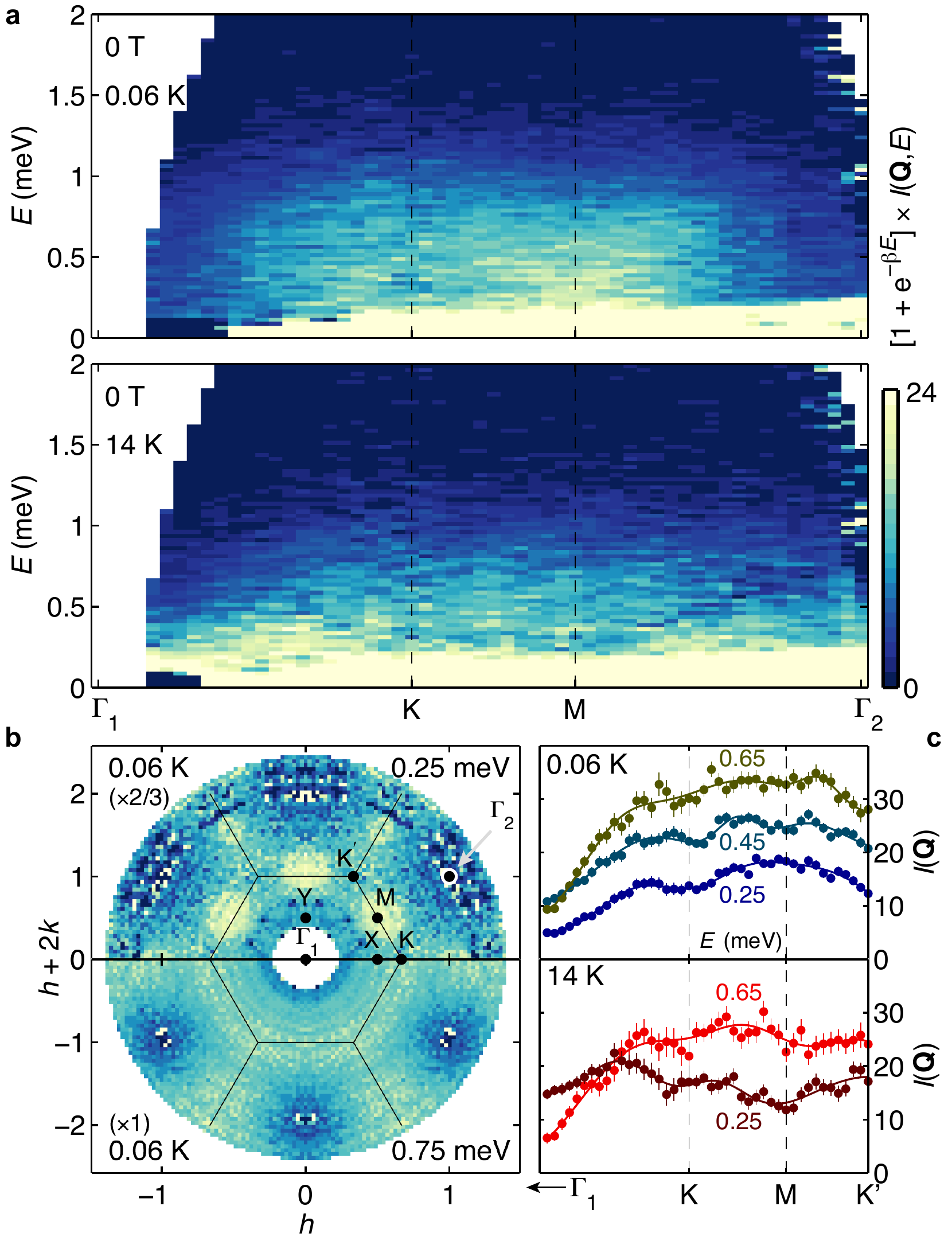}
	\end{center}
		\caption{\label{fig2} {\bf Neutron-scattering data for YbMgGaO{$_4$} measured in zero applied field.} {\bf a.} Energy dependence of magnetic  excitations along high-symmetry directions in reciprocal space, showing data at 0.06\,K (upper panel) and 14\,K (lower panel). Reciprocal-space points are labeled in {\bf b} and scattering intensity in arbitrary units is shown as a color scale. \red{The temperature of 0.06 K was measured at the mixing chamber of our dilution refrigerator.} {\bf b.} Wave-vector dependence of magnetic excitations at 0.06\,K, at energy transfers of 0.25\,meV (upper panel) and 0.75\,meV (lower panel).  {\bf c.} One-dimensional cuts along high-symmetry directions at 0.06\,K (upper panel) and 14\,K (lower panel). Each panel shows energy transfers of 0.25, 0.45, and 0.65\,meV (labeled on the graph) shifted vertically for clarity. \red{For {\bf a} and {\bf c}, the incident neutron energy $E_\mathrm{i}=3.32$\,meV and the data have been integrated over the vertical range $-\frac{1}{2} \leq l \leq \frac{1}{2}$; for {\bf b}, $E_\mathrm{i}=12$\,meV and $-1 \leq l \leq 1$. The scattering intensity in the top panel of {\bf b} is multiplied by a factor $2/3$ to share the same color scale as {\bf a}.}}
\end{figure}
%================  ================  ================  ================

To enable such experiments, we have grown a large single crystal of YbMgGaO{$_4$} using the floating-zone technique [Fig.~S1]. We characterized our crystal using specific heat, magnetization, and single-crystal neutron-diffraction measurements. In addition, we compared specific-heat and X-ray diffraction (XRD) measurements of crushed crystal pieces with the same measurements of a powder sample prepared by the solid-state method of \cite{Li_2015a}. Single-crystal neutron-diffraction measurements [Fig.~S2] reveal that our single crystal is predominantly a single grain and is therefore suitable for inelastic neutron-scattering measurements. The magnetic specific heat shows a broad peak at $T\approx3$\,K in zero magnetic field, and a $T^{\gamma}$ dependence with $\gamma=0.703(4)$ below the peak [Fig. 1{\bf b}], consistent with previous experimental results \cite{Li_2015a}. Our magnetic entropy [Fig. 1{\bf c}] and magnetization [Fig. 1{\bf d}] measurements fully support the effective spin-$1/2$ scenario (with $g_{\parallel}\approx3.7$) for Yb$^{3+}$ at low temperatures. Rietveld refinements to our XRD data [Fig.~S3] yield a good fit for the published structural model \cite{Li_2015a} with no Yb$^{3+}$/Ga$^{3+}$ site mixing observed within our experimental accuracy [Tab.~S1]. We observe no significant differences between powder and crushed single-crystal samples, indicating that the sample dependence that has afflicted QSL candidates such as Yb$_2$Ti$_2$O$_7$ \cite{Yaouanc_2011} is not evident in YbMgGaO{$_4$}.

Neutron-scattering experiments measure spin correlations as a function of energy and reciprocal space, yielding direct information about correlated quantum states \cite{Marshall_1968}. Single-crystal neutron-scattering data measured in zero field are shown in Fig.~2, and provide strong evidence for QSL behavior. Throughout, we plot neutron-scattering intensity as $(1+e^{-\beta E})\,I(\mathbf{Q},E)$, where $I(\mathbf{Q},E)$ is the measured intensity as a function of scattering wave-vector $\mathbf{Q}=h\boldsymbol{a}^\ast + k\boldsymbol{b}^\ast + l\boldsymbol{c}^\ast$ and energy transfer $E$, and the factor $1+e^{-\beta E}$ corrects for detailed balance \cite{Marshall_1968}. Fig.~2{\bf a} shows the energy dependence of the scattering intensity along high-symmetry reciprocal-space directions. The scattering at $E\gtrsim 0.2$\,meV is magnetic, as shown by its dependence on applied magnetic field (discussed below). At both 0.06\,K and 14\,K, our data reveal a broad continuum of excitations, \red{in contrast to the spin-wave scattering observed in the spin-1/2 triangular-lattice compound Ba$_3$CoSb$_2$O$_9$~\cite{Ma_2016}}. The presence of an energy continuum shows that YbMgGaO{$_4$} does not exhibit spin freezing at 0.06\,K, and is consistent with the fractionalized (spinon-like) excitation spectrum expected in a QSL \cite{Kalmeyer_1987,Han_2012}. The excitation continuum \red{has a bandwidth of 1.3 meV and} is gapless within the experimental energy resolution of approximately $0.1$\,meV. \red{This scattering spans a much smaller energy scale but is otherwise qualitatively similar to} herbertsmithite \cite{Vries_2009,Han_2012}---a surprising result, because the kagome lattice of herbertsmithite \red{is considered} more highly frustrated than the triangular lattice of YbMgGaO{$_4$} \cite{Capriotti_1999,Yan_2011}.

The wave-vector dependence of the 0.06\,K scattering intensity is shown in Fig.~2{\bf b}. The elastic scattering is much broader than the instrumental resolution ($\delta Q/Q \approx 2\%$) and reveals no magnetic Bragg peaks, confirming the absence of magnetic order at 0.06\,K. At low energy (0.25\,meV), the inelastic scattering shows a broad maximum at the $\mathrm{M}\equiv(\frac{1}{2}00)$ point of the hexagonal Brillouin zone, whereas at higher energy (0.75\,meV) the scattering is isotropic around the zone boundary. Further insight into this feature is obtained from line plots across the M-point maximum at different energies [Fig.~2{\bf c}]. At 0.06\,K, these plots reveal a weak dispersive excitation \red{originating from the} $\mathrm{M}$ point, superimposed on the continuum shown in Fig.~2{\bf a}. Strikingly, \red{some structure persists} in the excitation spectrum at 14\,K ($\gtrsim 3\theta_\mathrm{W}$), but the location of the intensity maximum at low energy ($\sim$\,0.25\,meV) shifts to between $\Gamma\equiv(000)$ and $\mathrm{K}\equiv(\frac{1}{3}\frac{1}{3}0)$ points [Fig.~2{\bf c}]. These features are reminiscent of spinon excitations in the antiferromagnetic Heisenberg spin chain \cite{Starykh_1997,Mourigal_2013}---the prototypical example of a one-dimensional QSL state---but have not previously been observed in two-dimensional QSL candidates. 

%================  ================  ================  ================
\begin{figure*}
	\begin{center}
		\includegraphics[scale=0.55]{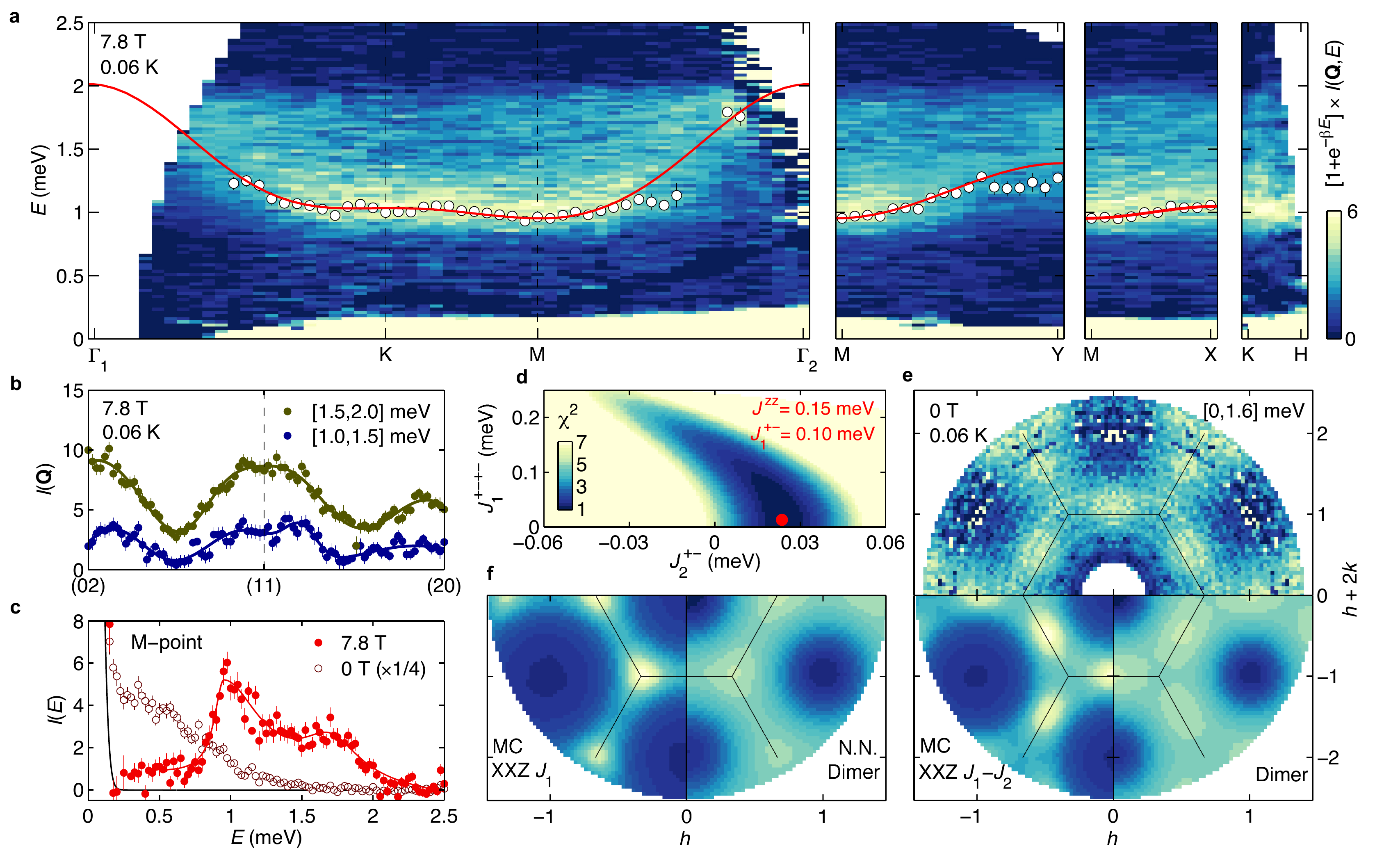}
	\end{center}
		\caption{\label{fig3} {\bf Field-polarized neutron-scattering data and evidence for next-nearest-neighbor interactions in YbMgGaO{$_4$}}. {\bf a.} Energy dependence of magnetic excitations along high-symmetry directions, measured at 0.06\,K in an applied field of 7.8\,T. The white circles show the location of the maximum intensity at each wave-vector (see text). The red lines show a fit to the spin-wave dispersion relation, Eq.~(2). The labeling of reciprocal-space positions is given in Fig.~2{\bf b}. The rightmost panel, for which \red{H=$(\frac{1}{3}\frac{1}{3}\frac{1}{2})$}, demonstrates the absence of dispersion along $l$ (perpendicular to the triangular planes). {\bf b.} One-dimensional plots of the 0.06\,K magnetic intensity, showing that the scattering peaks \red{around} the zone centers when a 7.8\,T field is applied. The range of $E$-integration is labeled on the graph. {\bf c.} Representative fit to the $E$-dependence of the data shown in {\bf a}, used to determine the position of the intensity maximum at a given wave-vector [white circles in (a)]. {\bf d.} Dependence of the goodness-of-fit parameter $\chi^2$ on the values of the exchange interactions $J_{1}^{\pm\pm}$ and $J_{2}^{\pm}$. {\bf e.} Experimental diffuse-scattering data at 0.06\,K obtained by integrating over energy transfer (top panel); calculation from classical Monte Carlo simulations of our model with nearest and next-nearest neighbor interactions (bottom left panel); and calculation from a phenomenological model of nearest-neighbor and next-nearest-neighbor dimers (bottom right panel). {\bf f.} Calculated diffuse scattering for models with nearest-neighbor interactions only, showing Monte Carlo calculation (left panel) and dimer model (right panel). The data in {\bf a} and {\bf c} were measured with $E_\mathrm{i}=3.32$\,meV, and the data in {\bf b} and {\bf e} with $E_\mathrm{i}=12$\,meV.}
\end{figure*}
%================  ================  ================  ================

What is the probable origin of the QSL state in YbMgGaO{$_4$}? To answer this question, we measured the excitation spectrum with a large magnetic field applied along the $c$ axis. These data are information-rich, because they allow exchange constants to be fitted to the dispersion curves of the field-polarized state by applying linear spin-wave theory \cite{Ross_2011}. The starting point for our analysis is the effective spin-$1/2$ Hamiltonian \cite{Li_2015,Ross_2011,Onoda_2011},
\begin{eqnarray}\label{eq:Hamiltonian}
	\mathcal{H} & = & ~~~~\sum_{\left\langle i,j\right\rangle }\big[J_{1}^{zz}S_{i}^{z}S_{j}^{z}+J_{1}^{\pm}\left(S_{i}^{+}S_{j}^{-}+S_{i}^{-}S_{j}^{+}\right)
	 + J_{1}^{\pm\pm}\left(\gamma_{ij}S_{i}^{+}S_{j}^{+}+\gamma_{ij}^{*}S_{i}^{-}S_{j}^{-}\right)\big]  \nonumber \\
	 &  & +\sum_{\left\langle \left\langle i,j\right\rangle \right\rangle }\big[J_{2}^{zz}S_{i}^{z}S_{j}^{z}+J_{2}^{\pm}\left(S_{i}^{+}S_{j}^{-}+S_{i}^{-}S_{j}^{+}\right)\big]
	 -\mu_{0}\mu_{\mathrm{B}}\sum_{i}\big[g_{\perp}\left(H^{x}S_{i}^{x}+H^{y}S_{i}^{y}\right)+g_{\parallel}H^{z}S_{i}^{z}\big],
\end{eqnarray}
where spin operators $S^{\pm} = S^{x} \pm \mathrm{i}S^{y}$, $g_{\parallel}$ and $g_{\perp}$ are the components of the $g$-tensor parallel and perpendicular to the applied field, and the complex numbers $\gamma_{ij}$ are defined in \cite{Li_2015}. Angle brackets $\left\langle ~ \right\rangle $ and $\left\langle \left\langle ~ \right\rangle \right\rangle $ indicate that the sum is taken over all nearest and next-nearest neighbor pairs, respectively (with each pair counted once). The exchange interactions $J^{zz}$ and $J^{\pm}$ define an XXZ model \cite{Yamamoto_2014}, where we include nearest and next-nearest neighbor interactions (denoted by subscripts ``1" and ``2", respectively). We also include a symmetry-allowed bond-dependent interaction, $J_{1}^{\pm\pm}$; however, we neglect the other bond-dependent interaction ($J_{1}^{z\pm}$ in \cite{Li_2015}), because our experimental data are insensitive to this term and ESR measurements indicate that takes a very small value \cite{Li_2015}. For a large applied field along $c$, it follows from Eq.~(1) that the spin-wave dispersion is given by \cite{Li_2015b}
\begin{eqnarray}\label{eq:dispersion}
	[\varepsilon(\mathbf{Q})]^2 & = & \biggl\{ g_{\parallel}\mu_{\mathrm{B}}\mu_{0}H^{z}-3\left(J_{1}^{zz}+J_{2}^{zz}\right)
	 +\sum_{i=1}^{6}\left[J_{1}^{\pm}\cos\left(\mathbf{Q}\cdot\mathbf{r}_{1,i}\right)+J_{2}^{\pm}\cos\left(\mathbf{Q}\cdot\mathbf{r}_{2,i}\right)\right]\biggr\}^{2} \nonumber \\
	 &  & -\left|J_{1}^{\pm\pm}\sum_{i=1}^{6}\gamma^{*}_{1,i}\cos\left(\mathbf{Q}\cdot\mathbf{r}_{1,i}\right)\right|^{2},
\end{eqnarray}
where $\mathbf{r}_{1,i}$ and $\mathbf{r}_{2,i}$ label the nearest and next-nearest neighbor vectors, respectively.

Our experimental data measured in a 7.8\,T field close to the $\approx$\,8\,T saturation are shown in Fig.~3{\bf a}. As anticipated from Eq.~(2) and from the field dependence of the specific heat [Fig.~1{\bf b}], the applied field induces a spin gap of \red{$1.0$~meV, and a single dispersive signal with a bandwidth of $1.1$~meV} dominates the spectrum. A continuum of excitations is also visible, indicating that the applied field may not be strong enough to suppress longitudinal spin fluctuations entirely. Nevertheless, the inelastic scattering is peaked very close to the zone centers, demonstrating that the system is essentially polarized [Fig.~3{\bf b}]. \red{The intensity of the inelastic scattering drops significantly from 0 to 7.8\,T and exposes a redistribution of spectral weight: fully fluctuating spins in zero field evolve into a field-induced ferromagnet with non-zero uniform magnetization (and thus elastic scattering) and spin fluctuations constrained perpendicular to the applied field~\cite{Mourigal_2013}.} 

To model the dispersion, we first fit the $E$-dependence at each $\mathbf{Q}$ with a double \red{(i.e. asymmetric)} Lorentzian to determine the peak position; a representative fit is shown in Fig.~3{\bf c}. We then fit Eq.~(2) to the extracted dispersion curve by varying the exchange parameters $J_{1}^{zz}+J_{2}^{zz},J_{1}^{\pm}$, and $J_{2}^{\pm}$. \red{The choice of a purely 2D model is justified by the absence of visible dispersion along $l$ [Fig.~3({\bf a})].}  Throughout, we fix $g_{\parallel}=3.721$ from magnetization data \cite{Li_2015}. Initially, we also fix $J_{1}^{\pm\pm}=0.013$\,meV  from ESR measurements \cite{Li_2015}. We obtain an excellent fit with $J_{1}^{zz}+J_{2}^{zz}=0.154(3)$, $J_{1}^{\pm}=0.109(4)$, and $J_{2}^{\pm}=0.024(1)$\,meV. The ratio $J^{zz}/J^{\pm}=1.16(4)$ is comparable to the result from magnetization measurements ($J^{zz}/J^{\pm}=1.09(13)$ \cite{Li_2015}), locating the interactions of YbMgGaO{$_4$} between isotropic and planar limits ($J^{zz}/J^{\pm}$ equal to $2$ and $0$, respectively). In the limit of planar nearest-neighbor interactions, a gapless Dirac QSL may be stabilized \cite{Alicea_2005}; however, this phase does not persist for $J_{1}^{zz}/J_{1}^{\pm}>0$ if only nearest-neighbor interactions are present \cite{Yamamoto_2014}.
Of particular interest are the interactions $J_{1}^{\pm\pm}$ and $J_{2}^{\pm}$, because both may in principle stabilize a QSL \cite{Iqbal_2016,Li_2015b}. Fig.~3{\bf d} shows the dependence of the goodness-of-fit on the values of these parameters. Our dispersion curve is relatively insensitive to $J_{1}^{\pm\pm}$: the best fit is for $J_{1}^{\pm\pm}=0$, but the previously-reported value of $0.013$\,meV \cite{Li_2015} yields visually indistinguishable results. Crucially, however, our data are highly sensitive to $J_{2}^{\pm}$. To match the shallow dispersion minimum at the M point requires $J_{2}^{\pm}/J_{1}^{\pm}\approx 0.22$: smaller values of $J_{2}^{\pm}$ yield a minimum at the K point, while larger values yield a minimum at the M point that is too deep. Remarkably, this ratio lies close to the QSL regime of the \red{spin-1/2} $J_{1}$-$J_{2}$ Heisenberg triangular-lattice antiferromagnet, which is predicted to occur for $0.06 \lesssim J_{2}/J_{1} \lesssim 0.19$ \cite{Manuel_1999,Li_2015c,Zhu_2015,Hu_2015,Iqbal_2016}. \red{This points towards an important role of} antiferromagnetic $J_{2}^{\pm}$ in stabilizing a QSL state in YbMgGaO{$_4$}.

The magnetic diffuse scattering provides an independent check on our results, because it is sensitive to the instantaneous spin-pair correlations in zero applied field. We obtain the experimental diffuse intensity $I(\mathbf{Q})=\int_{0}^{E^{\prime}}(1+e^{-\beta E})I(\mathbf{Q},E)\thinspace\mathrm{d}E$, where $E^{\prime}=1.6$\,meV, and compare these data with calculations in Fig.~3{\bf e}-{\bf f}. At the simplest level, our data can be understood in terms of a spin-dimer model: \red{a model of nearest-neighbor dimers fails to match the data; however, including 20\% of next-nearest-neighbor dimers yields much better agreement, resembling observations on the triangular-lattice cluster-magnet LiZn$_2$Mo$_3$O$_8$~\cite{Mourigal_2014}}. \red{A Lorentzian fit along the $\Gamma$---M---$\Gamma$ path shows that correlations are comparable to the nearest-neighbor distance between Yb$^{3+}$ atoms, yielding an exponential correlation length of 2.8(8)\,\AA.} We back up our results with classical Monte Carlo simulations, driven by Eq.~(1) with our fitted values of the exchange interactions (we assume an XXZ model with $J_{1}^{zz}/J_{1}^{\pm}=J_{2}^{zz}/J_{2}^{\pm}$, and $J_{1}^{\pm\pm}=0$; including nonzero $J_{1}^{\pm\pm}$ yields additional features not observed in our data, as shown in Fig.~S4). At a simulation temperature of 1.3\,K, the calculated diffuse intensity for $J_{2}^{\pm}=0$ has its maximum at the K point, contrary to experiment. However, including $J_{2}^{\pm}=0.22J_{1}^{\pm}$ reproduces the M-point maximum observed experimentally [Fig.~3{\bf e-f}], providing evidence for next-nearest-neighbor interactions in zero field. Quantum calculations show the same trend towards M-point scattering with increasing $J_2$ \cite{Iqbal_2016}, in agreement with our results. \red{Crucially, after cooling our Monte Carlo simulations below 1.3\,K, the calculated diffuse scattering shows much sharper features than the experimental data [Fig.~S5]. Quantum fluctuations are a natural contender to explain the suppressed spin correlation length we observe at 0.06\,K.}

\red{The observation of an extended QSL regime in YbMgGaO{$_4$} sets it apart from other inorganic triangular-lattice magnets with quantum spins, in which magnetic order or spin freezing typically occurs at a temperature $\sim$\,$\theta_\mathrm{W}/10$ (see, e.g., \cite{Tokiwa_2006,Doi_2004,Nakatsuji_2005,Ma_2016}). In contrast, YbMgGaO{$_4$} shows a continuous excitations spectrum without symmetry-breaking magnetic order to temperatures below $\sim$\,$\theta_\mathrm{W}/70$.} Our results identify YbMgGaO{$_4$} as a QSL in which the magnetic scattering peaks at the M point of the triangular-lattice Brillouin zone. In this respect, \red{spins in YbMgGaO{$_4$} appear more correlated than for} herbertsmithite---in which the magnetic scattering is essentially isotropic along the zone boundaries \cite{Han_2012}---and for the nearest-neighbor resonating valence-bond model \cite{Anderson_1973}. The balance between planar anisotropy and next-nearest-neighbor interactions determined from our experiments provides a reference-point for theoretical investigations of the stability and classification of QSL states \cite{Wen_2002}. \red{Furthermore, our parameters should allow theoretical studies to determine if YbMgGaO$_4$ lies in proximity to a quantum critical point towards a magnetically-ordered state}. Compared to organic triangular QSL candidates \cite{Shimizu_2003,Pratt_2011,Yamashita_2011}, the strength of the magnetic signal and the availability of large single-crystal samples makes YbMgGaO{$_4$} an exceptional candidate for neutron-scattering experiments, suggesting that mapping the response of a \red{2D} QSL to temperature and applied magnetic field is now a practical prospect.

%======================================================================
% Main Bibliography
%======================================================================
\subsection*{References}
\begingroup
\renewcommand{\section}[2]{}%

\endgroup

%======================================================================
% Methods
%======================================================================
\subsection*{Methods}

\noindent\textit{Sample preparation.} 
A polycrystalline sample of YbMgGaO$_4$ was synthesized by a solid state method. Stoichiometric ratios of Yb$_2$O$_3$, MgO, and Ga$_2$O$_3$ fine powder were carefully ground and reacted at a temperature of 1450$^\circ$\,C for 3 days with several intermediate grindings. The single-crystal sample of YbMgGaO$_4$ [Fig.~S1] was grown using the optical floating-zone method under a 5\,atm oxygen atmosphere~\cite{Li_2015}. The best single crystal was obtained with a pulling speed of 1.5\,mm/h, and showed cliffed $[001]$ surfaces after several hours of growth. \\

\noindent\textit{X-ray diffraction measurements and refinements.} 
Room-temperature powder X-ray diffraction (XRD) were carried out on powder and crushed single-crystal samples using a Panalytical X'pert Pro Alpha-1 diffractometer with monochromatic Cu-$K\alpha$ radiation ($\lambda = 1.540598$\,\AA). Preliminary measurements in flat-plate geometry on a loose powder using a Huber X-ray diffractometer showed preferred orientation, especially for the crushed-crystal sample. To minimize the extent of preferred orientation, we loaded our samples into a glass capillary that was rotated at 30\,rpm. Due to the large absorption cross-section of Yb, these measurements were limited to small sample sizes. Measurements were taken between $5\leq 2\theta \leq 140^\circ$ with $\Delta2\theta = 0.016^\circ$. Rietveld refinement was carried out using the FULLPROF program \cite{Rodriguez-Carvajal_1993a}.   
Peak-shapes were modelled by pseudo-Voigt functions, and the remaining preferred orientation was treated within the March model \cite{Dollase_1986}. Fits to data are shown in Fig.~S3, and refined values of structural parameters are given in Tab.~S1.\\

\noindent\textit{Thermomagnetic measurements.} 
Heat-capacity measurements were carried out on a Quantum Design PPMS instrument using dilution fridge ($0.06 \leq T\leq 4$\,K) and standard ($1.6 \leq T \leq 100$\,K) probes in a range of measuring magnetic fields, $0 \leq \mu_0 H \leq 14$\,T. Single-crystal measurements were made on a flat thin piece polished to $\approx$\,1\,mg and oriented with the $c$ axis parallel to the applied magnetic field. To ensure sample thermalization at low temperatures, powder measurements were made on pellets of YbMgGaO$_4$ mixed with an approximately equal mass of silver powder, the contribution of which was measured separately and subtracted to obtain the specific heat $C_p$. The magnetic specific heat $C_\mathrm{m}$ was obtained by subtracting a modelled lattice contribution $C_\mathrm{L}$ with two Debye temperatures, $480$\,K and $142$\,K. The change in magnetic entropy $\Delta S(T)$ was subsequently obtained by integrating $C_\mathrm{m}/T$ from 0.06~K to $T$. Isothermal magnetization measurments $M(H)$ were performed on the above single-crystal piece using a PPMS vibrating sample magnetometer in a range of magnetic fields $0 \leq \mu_0 H \leq 14$\,T and temperatures $1.7 \leq T\leq 10$\,K. \\

\noindent\textit{Neutron scattering measurements.} 
Inelastic neutron-scattering experiments were performed on the CNCS spectrometer at the Spallation Neutron Source (SNS), Oak Ridge National Laboratory, USA \cite{Ehlers_2011}. The sample was a $2.2$\,g rod-shaped crystal cut into two shorter pieces to fit in the bore of a cryomagnet. The two pieces (total dimensions $16 \times 16\times 4 $\,mm$^3$) were co-aligned to within 1.5$^{\circ}$ using a Multiwire X-ray Laue backscattering machine, and mounted in the $(hk0)$ scattering plane on a oxygen-free copper holder [Fig.~S1]. The mount was attached to the bottom of a dilution refrigerator reaching a base temperature of 0.06\,K at the mixing chamber, indicating a sample temperature $\lesssim 0.1$\,K. The sample stick was inserted in an 8\,T superconducting cryomagnet, and measurements were performed in zero field and in a field of 7.8\,T applied along the $c$ axis. Two incident neutron energies were used, $E_\mathrm{i}=3.32$ and $12.0$\,meV, yielding elastic energy resolutions (FWHM) of 0.11 and 0.75\,meV, respectively. The sample was rotated in steps of 1$^\circ$, with a range of 270$^\circ$ for $E_\mathrm{i}=12$\,meV and 180$^\circ$ for $E_\mathrm{i}=3.32$\,meV. For $E_\mathrm{i}=3.32$\,meV, the background scattering from the sample environment was measured and subtracted from the data. For $E_\mathrm{i}=12$\,meV, the background and non-magnetic scattering at low energy ($E\leq$\,0.9\,meV) was subtracted using the 7.8\,T measurement, taking advantage of the spin gap induced by applied field. 

Elastic neutron-scattering experiments were performed on the CORELLI spectrometer at ORNL's SNS. One of the two above crystal pieces was attached to a copper pin at the bottom of a $^3$He cryostat reaching a base temperature of $0.3$\,K. The sample was aligned in the $(h0l)$ scattering plane to asses crystal quality and stacking of the triangular-lattice planes along $c$. Neutron-absorbing Cd was used to shield the sample holder and an empty cryostat measurement was used to remove the background contribution. The sample was rotated in steps of 6$^\circ$ over a 360$^\circ$ range, and measurements were taken at temperatures of $0.3$, $4.0$, and $40$\,K. Elastic neutron-scattering data measured at 0.3\,K are shown in Fig.~S2. \\

\noindent\textit{Data analysis.} 
Initial data reduction was performed in MANTID~\cite{Arnold_2014} for both CNCS and CORELLI datasets. For the CNCS data, subsequent analysis was performed in HORACE~\cite{Ewings_2016} on a dedicated node within Georgia Tech's Partnership for Advanced Computing infrastructure. To increase counting statistics, inelastic-scattering data were symmetrized into an irreducible 60$^\circ$ wedge of the hexagonal reciprocal lattice. The CORELLI data was normalized to absolute units in MANTID~\cite{Michels-Clark_2016}. \\

\noindent\textit{Monte Carlo simulations.} To perform classical Monte Carlo simulations, we rewrite the spin
Hamiltonian, Eq.~(1), for zero field in terms of spin components $S^{x},S^{y},$ and $S^{z}$:
\begin{eqnarray*}
H & = & \sum_{\left\langle i,j\right\rangle }\biggl\{ J_{1}^{zz}S_{i}^{z}S_{j}^{z}+2J_{1}^{\pm}\left(S_{i}^{x}S_{j}^{x}+S_{i}^{y}S_{j}^{y}\right) +2J_{1}^{\pm\pm}\left[\left(S_{i}^{x}S_{j}^{x}-S_{i}^{y}S_{j}^{y}\right)\cos(\phi_{ij})-\left(S_{i}^{x}S_{j}^{y}+S_{i}^{y}S_{j}^{x}\right)\sin(\phi_{ij})\right]\biggr\}\\
 &  & +\sum_{\left\langle \left\langle i,j\right\rangle \right\rangle }\left[J_{1}^{zz}S_{i}^{z}S_{j}^{z}+2J_{1}^{\pm}\left(S_{i}^{x}S_{j}^{x}+S_{i}^{y}S_{j}^{y}\right)\right],
\end{eqnarray*}
where the phase factors $\phi_{ij}=\phi_{ji}=\left\{ 0,+\frac{2\pi}{3},-\frac{2\pi}{3}\right\} $
for nearest-neighbor bonds along the directions $\mathbf{a}$, $\mathbf{b}$,
and $\mathbf{a}+\mathbf{b}$, respectively (where $\mathbf{a}$ and
$\mathbf{b}$ are shown in Fig.~1{\bf a}). We keep $J_{1}^{zz}=0.126$ and $J_{1}^{\pm}=0.109$\,meV fixed throughout, with either  $J_{2}^{zz}=0.027$ and $J_{2}^{\pm}=0.024$\,meV or  $J_{2}^{zz}=0$ and $J_{2}^{\pm}=0$, as labelled in Fig.~3{\bf{e-f}}. In the main text, we take $J_{1}^{\pm\pm}=0$; the effect of nonzero $J_{1}^{\pm\pm}$ is discussed in SI. 
We take the length of spin vectors as $\sqrt{S(S+1)}=\sqrt{3}/2$. Simulations were performed on a three-dimensional spin configuration
consisting of nine triangular layers, each containing 504 spins. A proposed spin move consists of rotating a
spin by a small amount (chosen so that 50-70\% of proposed moves were accepted). The simulations were initialized
at a high temperature and cooled in gradual increments. At each temperature, the simulation was run for
at least $10t_{0}$ proposed moves, where $t_{0}$ is the number of proposed moves required to decorrelate
the system. The single-crystal diffuse scattering intensity was calculated as \cite{Squires_1978}
\[
I(\mathbf{Q})\propto\left[f(Q)\right]^{2}\left|\sum_{i}\mathbf{S}_{i}^{\perp}\exp\left(\mathrm{i}\mathbf{Q}\cdot\mathbf{r}_{i}\right)\right|^{2},
\]
where $f(Q)$ is the Yb$^{3+}$ magnetic form factor \cite{Brown_2004}, and $\mathbf{S}_{i}^{\perp}=\mathbf{S}_{i}-\mathbf{Q}(\mathbf{S}_{i}\cdot\mathbf{Q})/Q^{2}$
is the component of spin $\mathbf{S}_{i}$ perpendicular to $\mathbf{Q}$. The calculated pattern was slightly smoothed to allow calculation on an arbitrary $\mathbf{Q}$-grid. To increase statistical averaging, the calculated $I(\mathbf{Q})$ was averaged over 100 spin configurations and $\bar{3}m$ diffraction symmetry was applied.

\subsection*{Methods References}

\begingroup
\renewcommand{\section}[2]{}

\endgroup

%======================================================================
% Ancillary
%======================================================================

\subsection*{Acknowledgements}
We are very grateful to Luwei Ge for his help with heat-capacity measurements, Michael Waterbury and Yuan Wan for theory discussions, and John Carruth, Saad Elorfi, Michelle Everett, and Cory Fletcher for sample environment and instrument support during our neutron-scattering experiments. The work and equipment at the Georgia Institute of Technology (J.A.M.P., M. D. and M.M.) was supported by the College of Sciences and the Executive Vice-President for Research. The work at the University of Tennessee (Z.L.D. and H.D.Z.) was supported by the National Science Foundation through award DMR-1350002. The research at Oak Ridge National Laboratory's Spallation Neutron Source was sponsored by the U.S. Department of Energy, Office of Basic Energy Sciences, Scientific User Facilities Division.

\subsection*{Author contributions}
J.A.M.P., M.D., Z.D., G.E., Y.L., M.B.S., and M.M. performed neutron-scattering experiments. J.A.M.P., M.D., and M.M. analyzed the data. Z.L.D. and H.D.Z. made the sample. Z.L.D. and M.M. characterized the sample.  M.D. and M.M. aligned the sample. M.M. made the figures and J.A.M.P. wrote the paper with input from all authors. H.D.Z. and M.M. designed and supervised the project.

\subsection*{Additional information}
Correspondence and requests for materials should be addressed to M.M. (mourigal@gatech.edu).

\subsection*{Competing financial interests}
The authors declare no competing financial interests.

%======================================================================
% SI
%======================================================================

%==========Parameters of SI============
       \renewcommand\refname{References}
		% For section headers starting with S
       \renewcommand{\thesection}{S.\arabic{section}}
       \renewcommand{\thesubsection}{\thesection.\arabic{subsection}}
		% Hack for making SOM Equations Conform to Format e.g. (S1), (S2), etc
        \setcounter{equation}{0}
        \makeatletter %% With ams
        \def\tagform@#1{\maketag@@@{(S\ignorespaces#1\unskip\@@italiccorr)}}
        \makeatother
		% Hack for making figures Say \figurename S\thefigure, e.g. Figure S1:
        \setcounter{figure}{0}
        \makeatletter
        \makeatletter \renewcommand{\fnum@figure}
        {\figurename~S\thefigure}
        \makeatother
		% Hack for making tables Say \tablename S\thetable, e.g. Table S1:
        \setcounter{table}{0}
        \makeatletter
        \makeatletter \renewcommand{\fnum@table}
        {\tablename~S\thetable}
        \makeatother

%==========Title and Authors============
\title{
{\bf Continuous excitations of the triangular-lattice \\ quantum spin liquid YbMgGaO{$_4$}} \\ {\it Supplementary Information}}
\author{
 	Joseph A.~M. Paddison$^1$, Marcus Daum$^{1,\ddagger}$, Zhiling Dun$^{2,\ddagger}$, Georg Ehlers$^3$, \\ 
 	Yaohua Liu$^3$ , Matthew B. Stone$^3$, Haidong Zhou$^2$, Martin Mourigal$^{1,\ast}$ \\
	\normalsize{$^1$School of Physics, Georgia Institute of Technology, Atlanta, GA 30332, USA } \\
	\normalsize{$^2$Department of Physics and Astronomy, University of Tennessee, Knoxville, TN 37996, USA} \\
	\normalsize{$^3$Quantum Condensed Matter Division, Oak Ridge National Laboratory, Oak Ridge, TN 37831, USA} \\
	\\
	\normalsize{$^{\ddagger}$ These authors contributed equally to this work} \\
	\normalsize{$^{\ast}$ Email: mourigal@gatech.edu}
	}
\maketitle	
\tableofcontents
\clearpage
%================  ================  ================  ================

\clearpage

\section{Single-crystal mount for neutron-scattering measurements}
%================  ================  ================  ================
\begin{figure}[h!]
	\begin{center}
		\includegraphics[scale=0.4]{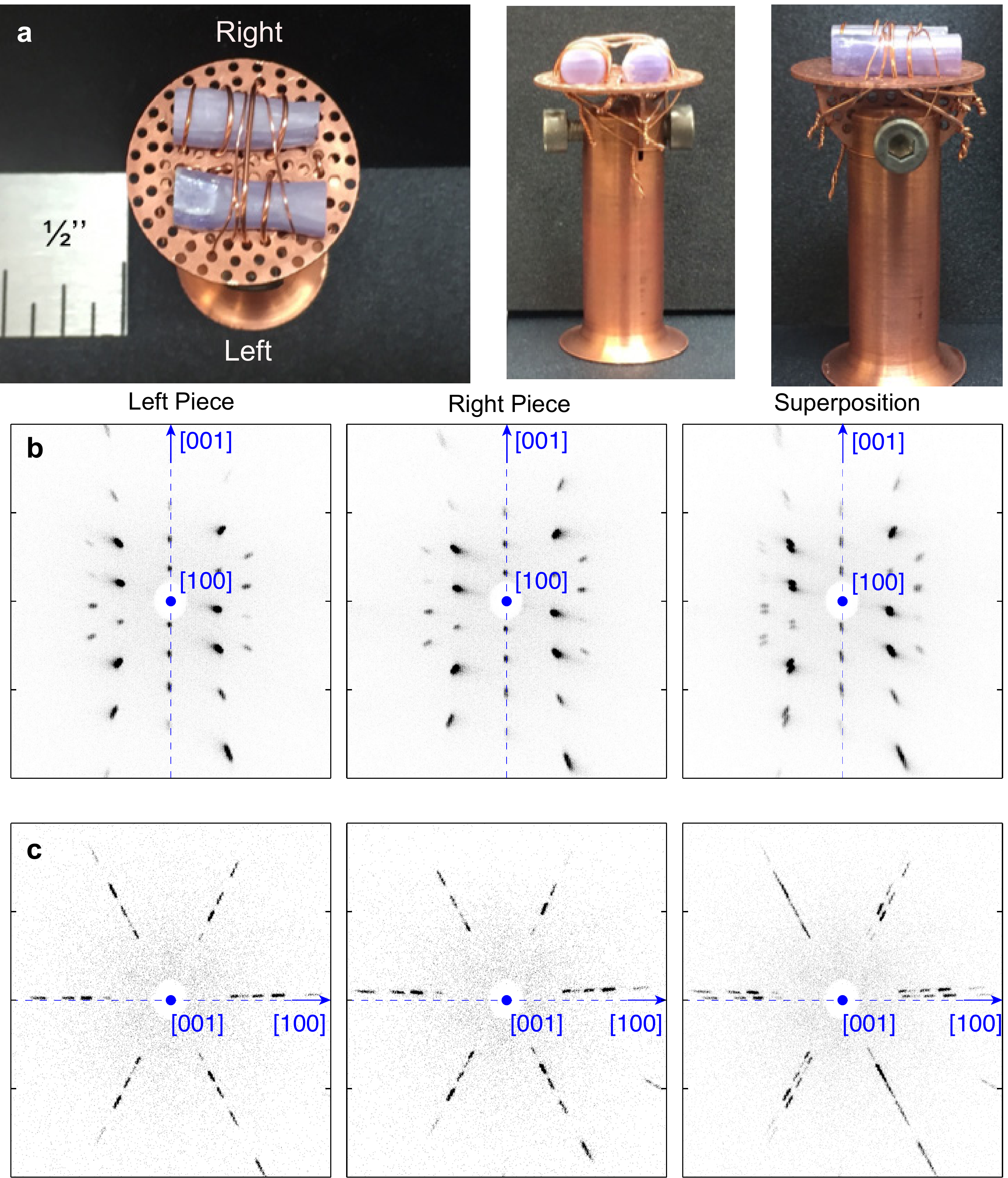}
	\end{center}
	\caption{{\bf a.} Photographs of the two co-aligned pieces of our single crystal, mounted in the $(hk0)$ scattering-plane on an oxygen-free copper sample holder. {\bf b.} Backscattering X-ray Laue pattern of the $(100)$ plane for each of the individual pieces (left and center plot) and co-alignment (right). {\bf c.} Backscattering X-ray Laue pattern of the $(001)$ plane for each of the individual pieces (left and center plot) and co-alignment (right).}
\end{figure}
%================  ================  ================  ================

\clearpage

\section{Elastic neutron-scattering measurements}
%================  ================  ================  ================
\begin{figure}[h!]
	\begin{center}
		\includegraphics[scale=0.7]{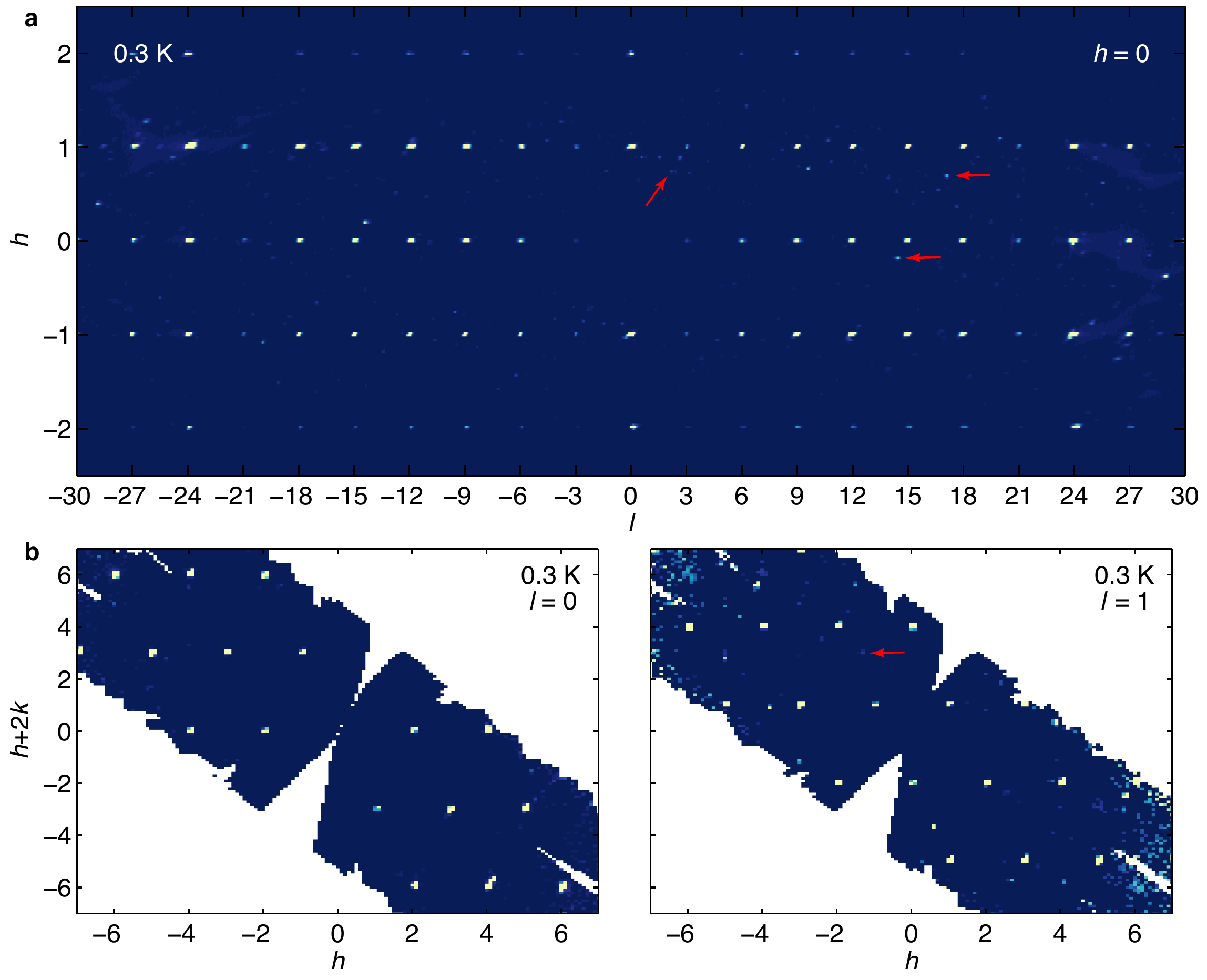}
	\end{center}
	\caption{Elastic neutron-scattering maps measured on one of our single-crystal pieces. The left panel shows the $(h0l)$ plane and the right panel shows the $(0kl)$ plane. The intensity scale is saturated at 0.1\% of the maximum intensity for each map. Red arrows indicate peaks arising from other crystal grains, which have a negligible effect for our inelastic neutron-scattering measurements because of their low intensity.}
\end{figure}
%================  ================  ================  ================

\clearpage

%======================================================================
\section{Powder X-ray diffraction measurements}

%================  ================  ================  ================
\begin{figure}[h!]
	\begin{center}
		\includegraphics[scale=1.0]{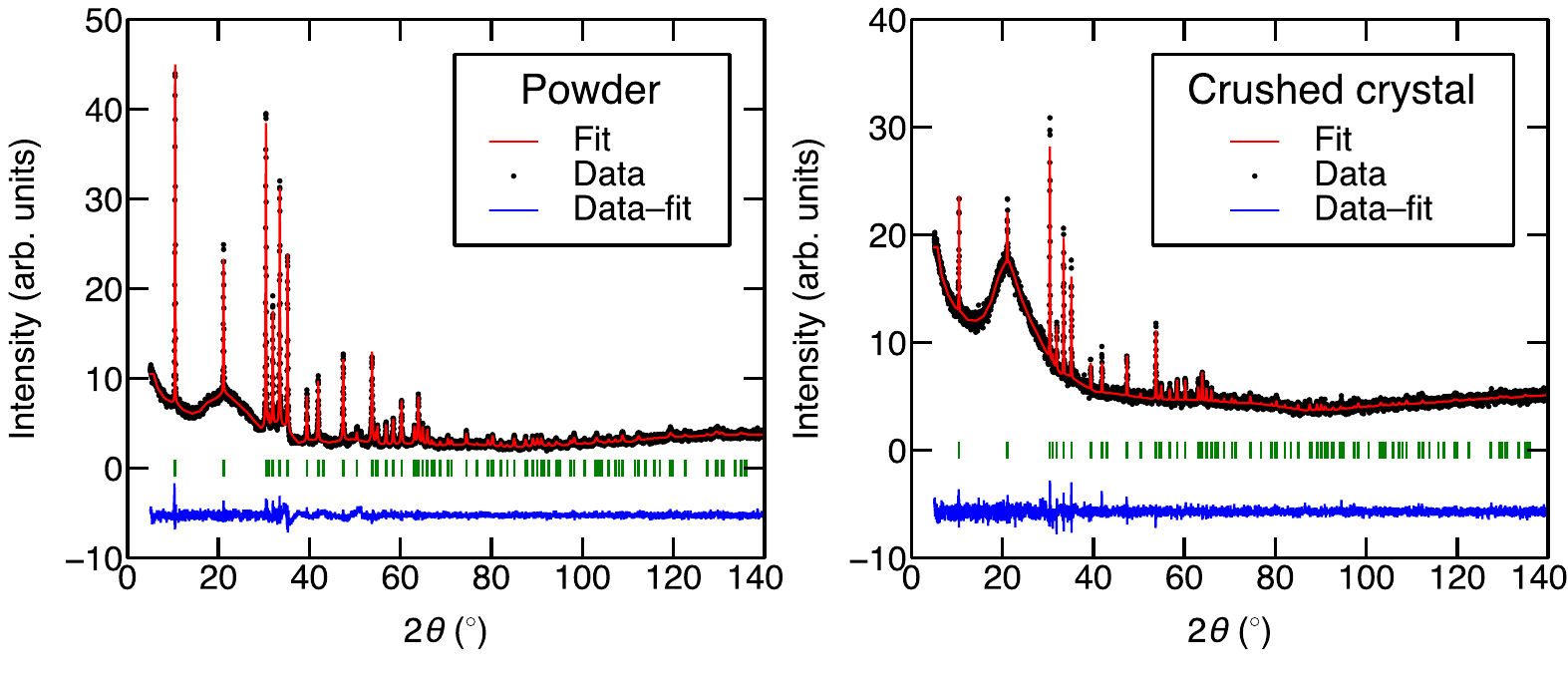}
	\end{center}
	\caption{Room-temperature X-ray diffraction patterns of powder (left) and crushed single-crystal (right), measured in a spinning glass capillary with $\lambda = 1.540598$\,\AA. The broad peak at $2\theta \sim 20^{\circ}$ is background from the capillary. The differences in relative peak intensities between powder and crushed-crystal datasets are explained by stronger preferred orientation for the crushed crystals [Tab.~S1].}
\end{figure}
%================  ================  ================  ================
\begin{table}[h!]
	\begin{centering}
	\begin{tabular}{cc|c||cc}
	\hline 
	\multicolumn{5}{c}{YbMgGaO$_4$, $R\bar{3}m$}\tabularnewline
	\hline 
	 &  & $T$ (K) & \multicolumn{2}{c}{300}\tabularnewline
	 &  & \multirow{1}{*}{Radiation} & \multicolumn{2}{c}{Cu $K_{\alpha}$, $\lambda=1.540598$\,\AA}\tabularnewline
	\hline 
	 &  &  & Powder & Crushed crystal\tabularnewline
	\hline 
	 &  & $a$ (\AA) & $3.4037(1)$ & $3.4018(1)$\tabularnewline
	 &  & $c$ (\AA) & $25.135(1)$ & $25.141(1)$\tabularnewline
	 &  & $p_{\mathrm{March}}$ & $1.026(3)$ & $1.189(7)$\tabularnewline
	 &  & \multirow{1}{*}{$R_{\mathrm{wp}}$ (\%)} & $6.26$ & $4.47$\tabularnewline
	\hline 
	Yb & $3a$, $(0,0,0)$ & $B_{\mathrm{iso}}$ (\AA$^{2}$) & $1.3(1)$ & $1.4(1)$\tabularnewline
	\hline 
	\multirow{2}{*}{Mg/Ga} & \multirow{2}{*}{$6c$, $(0,0,z)$} & $z$ & $0.2145(1)$ & $0.2143(4)$\tabularnewline
	 &  & $B_{\mathrm{iso}}$ (\AA$^{2}$) & $1.6(1)$ & $2.7(2)$\tabularnewline
	\hline 
	\multirow{2}{*}{O1} & \multirow{2}{*}{$6c$, $(0,0,z)$} & $z$ & $0.2896(5)$ & $0.2917(12)$\tabularnewline
	 &  & $B_{\mathrm{iso}}$ (\AA$^{2}$) & $1.7(2)^{\dagger}$ & $2.2(4)^{\ast}$\tabularnewline
	\hline 
	\multirow{2}{*}{O2} & \multirow{2}{*}{$6c$, $(0,0,z)$} & $z$ & $0.1293(4)$ & $0.1308(10)$\tabularnewline
	 &  & $B_{\mathrm{iso}}$ (\AA$^{2}$) & $1.7(2)^{\dagger}$ & $2.2(4)^{\ast}$\tabularnewline
	\hline 
	\end{tabular}
\par\end{centering}
\caption{Values of structural parameters determined from Rietveld refinement
against powder X-ray diffraction data. The parameter $p_\mathrm{March}$ is the March preferred-orientation parameter. The Mg/Ga site was constrained to be occupied by 50\% Mg and 50\% Ga, and pairs of parameters labelled with $^{\ast}$ and $^{\dagger}$ were constrained to be equal. We also tested a model of Ga$^{3+}$/Yb$^{3+}$ site mixing, which refined to zero (within uncertainty) for the powder, and to a negative (i.e., unphysical) value for the crushed crystal, without significantly improving the fit.}
\end{table}
%================  ================  ================  ================

\clearpage

\section{Effect of anisotropic exchange}

The simulations of the magnetic diffuse scattering shown in Fig.~3{\bf e-f} consider the XXZ model; i.e., the bond-dependent exchange interaction $J_{1}^{\pm\pm}=0$. Here, we show the effect of including nonzero $J_{1}^{\pm\pm}$ (as indicated by ESR measurements \cite{Li_2015}) on the diffuse scattering. First, we note that both ESR and our inelastic neutron-scattering measurements are sensitive only to the magnitude of $J_{1}^{\pm\pm}$. However, the sign of $J_{1}^{\pm\pm}$ strongly affects the diffuse scattering $I(\mathbf{Q})$ calculated from classical Monte Carlo simulations. Fig.~S4 shows calculated $I(\mathbf{Q})$  for $J_{1}^{\pm\pm}=\pm0.013$\,meV. In each case, we show the result for a model with nearest-neighbor interactions only, and for a model with both nearest and next-nearest-neighbor interactions. As noted in the text, including next-nearest-neighbor interactions shifts the maximum intensity to the M point, and this is unchanged by including small $J_{1}^{\pm\pm}$. However, introducing $J_{1}^{\pm\pm}$ also creates a modulation in scattering intensity, which depends on the sign of $J_{1}^{\pm\pm}$; this occurs because the spin configuration is no longer invariant to global rotations about $z$, and the diffuse scattering is sensitive the component of spin perpendicular to $\mathbf{Q}$ \cite{Squires_1978}. Evidently, including $J_{1}^{\pm\pm}=\pm0.013$\,meV yields less pleasing agreement with experimental data than was obtained for $J_{1}^{\pm\pm}=0$ [Fig.~3{\bf {e-f}}]. This apparent disagreement between Monte Carlo and ESR analysis \cite{Li_2015} may occur because our classical Monte Carlo simulations do not properly capture quantum effects associated with the anisotropic exchange interactions; alternatively, it may indicate that the line-widths observed in ESR \cite{Li_2015} have a different origin.

%================  ================  ================  ================
\begin{figure}[h!]
	\begin{center}
		\includegraphics[scale=1.0]{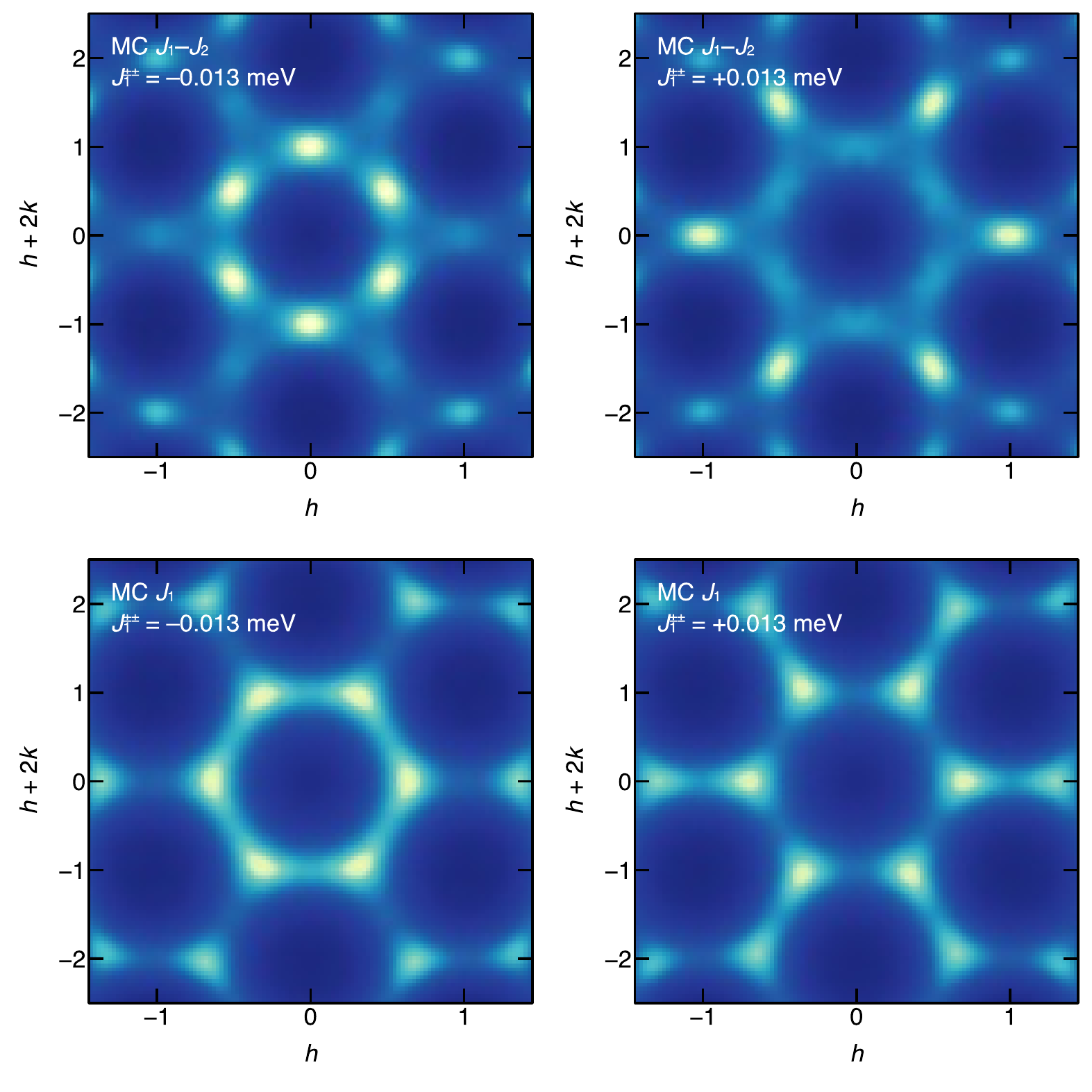}
	\end{center}
	\caption{Magnetic diffuse scattering calculated from classical Monte Carlo simulations including the bond-dependent anisotropic exchange interaction $J_{1}^{\pm\pm}$. For all calculations, we fix $J_{1}^{zz}=0.126$, $J_{1}^{\pm}=0.109$, and $|J_{1}^{\pm\pm}|=0.013$\,meV. 
	The top panels show calculations including nearest and next-nearest neighbor interactions, $J_{2}^{zz}=0.027$ and $J_{2}^{\pm}=0.024$\,meV, and the bottom panels show calculations including nearest-neighbor interactions only. The left panels show calculations for $J_{1}^{\pm\pm}=-0.013$\,meV, and the right panels show calculations for $J_{1}^{\pm\pm}=+0.013$\,meV. All calculations are at a temperature of 1.3\,K.}
\end{figure}
%================  ================  ================  ================

\clearpage

\section{Temperature dependence of magnetic diffuse scattering}
%================  ================  ================  ================
\begin{figure}[h!]
	\begin{center}
		\includegraphics[scale=1.0]{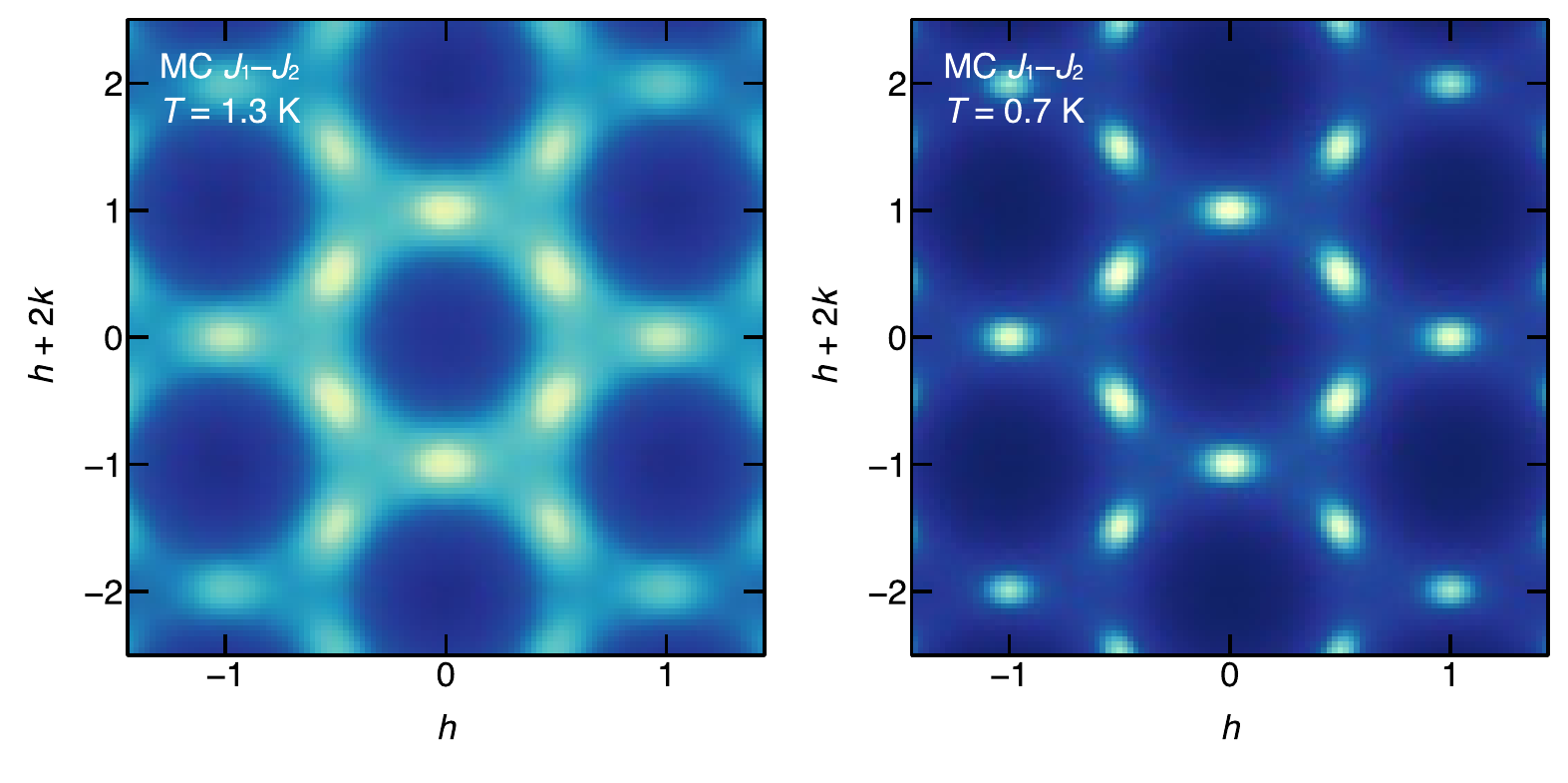}
	\end{center}
	\caption{Temperature dependence of magnetic diffuse scattering calculated from classical Monte Carlo simulations of the XXZ model with nearest and next-nearest neighbor interactions, showing 1.3\,K (left panel) and 0.7\,K (right panel). Intensities for the left panel are multiplied by 2 compared to the right panel. The exchange constants are $J_{1}^{zz}=0.126$, $J_{2}^{zz}=0.027$, $J_{1}^{\pm}=0.109$, and $J_{2}^{\pm}=0.024$\,meV, with $J_{1}^{\pm\pm}=0$.}
\end{figure}
%================  ================  ================  ================

%\section{Effect of anisotropic interactions on magnetic diffuse scattering}

%TBD

\end{document}